\documentclass[12pt,a4paper,british]{iopart}
\usepackage{amsfonts}
\usepackage{amssymb}
\usepackage{color}
\usepackage[T1]{fontenc}
\usepackage[latin9]{inputenc}
\usepackage{babel}
\usepackage{graphicx}
\usepackage{epstopdf}
\usepackage[numbers,square,sort&compress]{natbib}
\usepackage[unicode=true,pdfusetitle,bookmarks=false,breaklinks=false,pdfborder={0 0 1},backref=false,colorlinks=false]{hyperref}
\usepackage{array}
\usepackage{verbatim}
\usepackage{amstext}
\usepackage{iopams}
\usepackage{setstack}

\makeatletter

\newcommand{\bra}[1]{\langle #1|}
\newcommand{\ket}[1]{|#1\rangle}

\renewcommand{\t}[1]{\textrm{#1}}

\begin{document}

\title{True precision limits in quantum metrology}

\author{Marcin Jarzyna, Rafa{\l} Demkowicz-Dobrza{\'n}ski}
\address{Faculty of Physics, University of Warsaw, ul. Pasteura 5, PL-02-093 Warszawa, Poland}
\ead{Marcin.Jarzyna@fuw.edu.pl, demko@fuw.edu.pl}

\begin{abstract}
We show that quantification of the performance of quantum-enhanced measurement schemes
based on the concept of quantum Fisher information yields asymptotically equivalent results as the rigorous Bayesian approach,
provided generic uncorrelated noise is present in the setup.
At the same time, we show that for the problem of decoherence-free phase estimation
this equivalence breaks down and the achievable estimation uncertainty calculated within the Bayesian approach is by a $\pi$ factor larger than that predicted by the QFI even in the large prior knowledge (small parameter fluctuation) regime, where QFI is conventionally regarded as a reliable figure of merit.   We conjecture that the analogous discrepancy is present in arbitrary decoherence-free unitary parameter estimation scheme and propose a
general formula for the asymptotically achievable precision limit. We also discuss protocols utilizing states with indefinite number of particles and
show that within the Bayesian approach it is legitimate to replace the number of particles with the mean number of particles
in the formulas for the asymptotic precision, which as a consequence provides another argument that proposals based on the properties of the QFI of indefinite particle number states leading to sub-Heisenberg precisions are not practically feasible.
\end{abstract}


\maketitle

\section{Introduction}

Capability of performing precise measurements is the cornerstone of modern physics. Unlike classical physics, quantum mechanics provides
insight into fundamental limits on the achievable measurement precision that cannot be beaten irrespectively of the extent
of any future improvements in measurement technology. The paradigmatic example is that of the optical phase measurement.
Within the quantum optical framework the phase of a given state of light can only be defined up to a precision
that scales as $1/N$ where $N$ is the characteristic number of photons (proportional to mean energy) of a given state \cite{Zwierz2010,Giovanetti2012}.
This fact has profound implications on the performance of any metrological scheme based on optical interferometry where
the difference of optical phase delays in the respective arms of an interferometer is being sensed.
The achievable phase difference estimation precision is bounded by $\Delta\varphi \geq  1/N$ which is referred
to as the Heisenberg limit  \cite{Bollinger1996,Lee2002,Giovannetti2006}, as it may be informally viewed as a version of the
 Heisenberg uncertainty relation adapted to the phase-photon number case.
Presence of decoherence, however, which may be due to noise or experimental imperfections, typically prevents quantum-enhanced measurement schemes from reaching the Heisenberg scaling, and it may be demonstrated that for the generic uncorrelated noise processes
classically scaling bounds $\Delta \varphi \geq \t{const}/\sqrt{N}$ hold, limiting quantum enhancement to a constant factor precision improvement \cite{Huelga1997, Escher2011, Demkowicz2012}. Many of the bounds derived in the field of quantum metrology, including the ones mentioned above,
are applications of the celebrated Quantum Cram{\'e}r-Rao (C-R) bound  \cite{Helstrom1976, Holevo1982} which is based on calculation of the
Quantum Fisher Information (QFI).
The quantum C-R bound, however, may not be saturable in general, hence having derived the bounds it is highly relevant to ask whether there are explicit quantum estimation schemes that lead precisely to the minimal uncertainty predicted by the bounds.
This issue was particularly important in the recent discussion on possibility of reaching a sub-Heisenberg precision scaling, $\Delta \varphi = 1/N^\alpha$ with $\alpha >1$, where C-R based bounds apparently indicated such a possibility in certain setting involving indefinite particle number states \cite{Anisimov2010, Rivas2012, Zhang2013}, while at the same time it has been shown that the corresponding estimation scheme would require prior knowledge of the order of the value of the parameter estimation precision itself, limiting the practical usefulness of the proposals \cite{Giovanetti2012}.

The main goal of the present paper is to systematically investigate the problem of saturability of the quantum C-R based bounds
discussed in the quantum metrology literature. For this purpose we take the Bayesian approach to quantum estimation problems and
make a connection between the predictions of the Bayesian approach and that of the C-R bounds. Since the solution
to the Bayesian quantum estimation problem carries with itself an explicit description of the estimation protocol reaching the optimal precision it leaves no doubts on the issues of saturability. Admittedly, the Bayesian approach involves some degree of arbitrariness in defining the prior probability distribution describing the initial knowledge on the parameter value. However, for sufficiently regular priors their exact form
is not expected to affect the results valid in the asymptotic regime
of large resources, $N \rightarrow \infty$, and hence allow to draw conclusions on asymptotic scaling which are independent on the
form of the prior distributions.

Although C-R based approaches dominate the quantum metrology literature, there are also notable examples of a
successful application of the Bayesian paradigm. For example, it has been
demonstrated within the rigorous Bayesian estimation framework, that assuming a complete prior ignorance on the value of the estimated phase
one can at best reach $\Delta \varphi = \pi/N$ asymptotic precision  in decoherence-free phase estimation \cite{Berry2000}, which reveals a $\pi$ factor discrepancy between this result and the C-R based Heisenberg bound. On the other hand, if losses are taken into account the Bayesian
approach \cite{Kolodynski2010} yield the same asymptotic precision limit as predicted by the C-R based bounds \cite{Escher2011, Demkowicz2012}.

In this paper we put these observations into a wider context. We prove that in the decoherence-free phase estimation the asymptotic $\pi$ factor discrepancy is not due to a particular choice of prior distribution in the Bayesian setting, but holds also for arbitrary narrow regular priors.
We demonstrate this rigorously for Gaussian priors and support the conclusions with numerical evidence obtained for other priors as well.
Going beyond the phase estimation scheme, we also conjecture a general formula for the optimal precision of arbitrary decoherence-free unitary parameter estimation that is expected to be asymptotically saturable. More importantly, we show that in the presence of generic uncorrelated noise
which yields $\t{const}/\sqrt{N}$ precision scaling, the C-R bounds coincide with the precisions achievable by the
asymptotic Bayesian strategies and as such may be taken with full confidence. We finally prove that
the above conclusions apply also to metrological models involving indefinite number of particles where $N$ appearing in the formulas for
asymptotic precision may be confidently replaced with the mean number of particles $\bar{N}$ providing an alternative argument against the feasibility of sub-Heisenberg estimation procedures.

This paper is organized as follows. In section \ref{sec:QFI} we introduce local estimation approach based on the application of C-R bound and the use  of the QFI. In section \ref{sec:bayes} we describe Bayesian procedures which provide alternative way of defining precision. Section \ref{sec:decoherence} contains results about the asymptotic equivalence of both approaches in the presence of uncorrelated decoherence whereas in section \ref{sec:nodecoh} we discuss the differences that arises between them in the decoherence-free case. Section \ref{sec:global} contains an illustrative example to show that in some cases, such as in presence of correlated dephasing process, one cannot expect asymptotic irrelevance of the Bayesian prior and hence cannot compare the C-R and the Bayesian approaches in a meaningful way. Section \ref{sec:indefinite} contains a generalization of the obtained results to states with indefinite number of particles. Section \ref{sec:conclusions} summarizes the paper.

{\color{red}
\begin{figure}[t!]
\includegraphics[width=0.75\columnwidth]{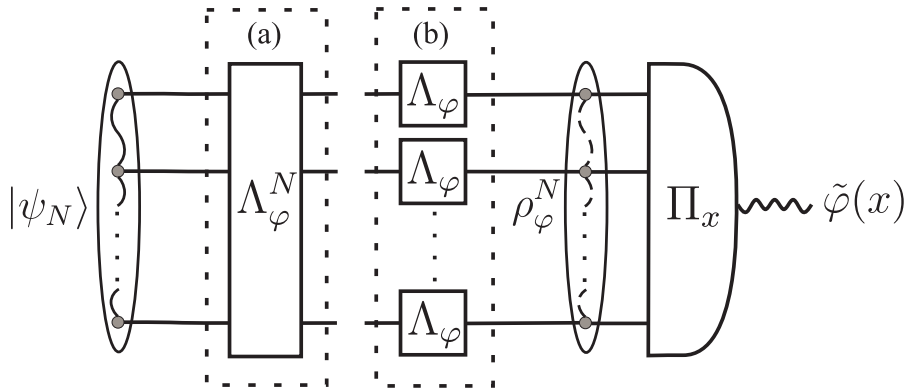}
\caption{Basic scheme of quantum metrology. $N$ particle state $\ket{\psi_N}$ is send through (a) general quantum channel (b) $N$ parallel quantum channels inscribing parameter value $\varphi$ as well as causing decoherence independently on each of the probes resulting in the output state $\rho^N_{\varphi}$. Measurement $\hat{\Pi}_x$ on the output state allows to make an estimate $\tilde{\varphi}(x)$ based on the measurement results $x$.}
\label{fig:scheme}
\end{figure}
}

\section{Cramer-Rao bound approach}
\label{sec:QFI}
For the purpose of this paper we consider a general estimation scheme  depicted in Fig.~\ref{fig:scheme} which is relevant for
 optical interferometry as well as more general quantum metrological protocols.
An $N$ particle probe state $\ket{\psi_N}$ undergoes an evolution described by the action of a quantum channel
$\Lambda^{N}_\varphi$. A general quantum measurement, $\{\Pi_x\}$, is performed on the output state $\rho^N_{\varphi} = \Lambda^{N}_\varphi(\ket{\psi_N}\bra{\psi_N})$ and based on the measurement
   result $x$ one estimates the value of an unknown parameter using an estimator function $\tilde{\varphi}(x)$.

The  main goal in quantum estimation theory is to find the minimal achievable estimation
  error $\Delta\varphi$ optimized over the choice of probe states, measurements and estimators.
  In general this is a very difficult task and therefore a lower bounds on estimation error are often considered instead of
  exact precision limits. The pursue to derive new useful quantum metrological lower bounds is an active field of research, see e.g. \cite{Tsang2012, Nair2012, Hall2012, Giovannetti2012}. Still, the bound most commonly used in the literature is the long-serving
  quantum C-R bound  \cite{Braunstein1994,Helstrom1976}
  \begin{equation}\label{eq:CR}
\Delta\varphi\geq\frac{1}{\sqrt{k F_\varphi}},\quad F_\varphi=\tr{\rho^N_\varphi L_\varphi^2},
\end{equation}
where $k$ is the number of independent repetitions of the experiment, $F_\varphi$ is the QFI while
$L_\varphi$ is the symmetric logarithmic derivative (SLD) operator defined implicitly by equation $\frac{d\rho^N_\varphi}{d\varphi}=\frac{1}{2}\left(\rho^N_\varphi L_\varphi+L_\varphi\rho^N_\varphi\right)$. It is noteworthy to emphasize that this bound is saturable under some general conditions which will be stated at the end of this section.
Deriving fundamental precision bounds using the QFI amounts to optimization over input probe states $\ket{\psi_N}$ that yield the maximal
$F_\varphi$. This task is relatively simple in case of decoherence-free
unitary parameter estimation models where $\Lambda^N_\varphi(\cdot) = U^{\otimes N}_\varphi \cdot  U^{\dagger \otimes N}_\varphi$,
with $U_\varphi = \exp(\mathrm{i} H \varphi)$ as the QFI may be related directly to the variance of the evolution generator $H$.
In this case the optimal C-R bound takes the form \cite{Giovannetti2006}:
\begin{equation}
\label{eq:crunitary}
\Delta \varphi \geq \frac{1}{N(\lambda_+ - \lambda_-)},
\end{equation}
with $\lambda_+$ and $\lambda_-$ being respectively the maximal and the minimal eigenvalues of $H$,
while the optimal input state is defined using the corresponding eigenstates
$\ket{\psi_N}= (\ket{+}^{\otimes N} + \ket{-}^{\otimes N})/\sqrt{2}$. In a special case of optical interferometry
where each photon is represented by a two-level system, with levels corresponding to the photon traveling
in one or the other arm of the interferometer, $H = \sigma_z/2$ and we recover the previously mentioned
Heisenberg bound $\Delta\varphi \geq 1/N$, whereas the optimal input probe state is the so-called N00N state
$\ket{\psi_{\t{N00N}}} = (\ket{0}^{\otimes N} + \ket{1}^{\otimes N})/\sqrt{2} = (\ket{N,0} + \ket{0,N})/\sqrt{2}$,
where the last form of the state is written in the mode occupation basis.
This result is usually contrasted with the precision achievable with uncorrelated input probes $\ket{\psi_N}= \ket{\psi}^{\otimes N}$, where the maximal
QFI scales linearly with $N$, $F_\varphi = N(\lambda_+ - \lambda_-)$, and hence bounds the achievable precision
by a $1/\sqrt{N}$ scaling formula, characteristic for classical estimation problems, where $N$ is a number of independent and identically distributed (i.i.d.) observations.

The situation is much more involved when decoherence is taken into account.
Even though there are various methods that allow to tackle the problem numerically with reasonable efficiency \cite{Demkowicz2009,Jarzyna2013,Macieszczak2013,Macieszczak2013a,Frowis2014}, any numerical approach breaks down in the asymptotic
 regime of large $N$.
Fortunately, in recent years, powerful analytical methods have been developed that allow to find
the maximal achievable QFI in the regime of large $N$ \cite{Escher2011, Demkowicz2012, Knysh2014}. These techniques allowed
to derive analytical precision bounds for a number of important models in quantum metrology, including lossy optical interferometry
and atomic interferometry in presence of dephasing, setting useful benchmarks for the whole field of quantum-enhanced metrology.
Of a particular interest are uncorrelated noise models, when $\Lambda^N_\varphi = \Lambda_\varphi^{\otimes N}$, see Fig.~\ref{fig:scheme}.
In this case it can been shown \cite{Fujiwara2008, Matsumoto2010, Demkowicz2012, Kolodynski2013}
that generically the asymptotic scaling of QFI is always linear $F \overset{N \rightarrow \infty}{=} \alpha  \cdot N$ and as such any quantum-enhanced benefits resulting from the use of entangled states are bounded by a constant factor gain over ,,classical'' protocols which utilize uncorrelated probes:
\begin{equation}
\label{eq:crdecoh}
\Delta\varphi \geq \frac{\t{const}}{\sqrt{N}}.
\end{equation}
Predictive power of the QFI bounds \eref{eq:crunitary} and \eref{eq:crdecoh} crucially depends on how tight they are and whether
they can in principle be saturated.
Since the bounds are obtained by maximization of the QFI over input states, this translates to the question of whether the QFI is indeed a proper
measure quantifying the performance of quantum-enhanced measurement protocols.

In principle the C-R bound, \eref{eq:CR}, can be saturated in the limit of many independent experiments, $k\to\infty$,
 by using the maximum likelihood estimator and performing the measurements in the eigenbasis of $L_\varphi$ \cite{Braunstein1994,Helstrom1976,BrandorffNielsen2000}.
Practical implications of this statement are far form obvious, however. The QFI is a point-estimation concept that depends
only on $\rho^N_\varphi$ and $\frac{d\rho_\varphi^N}{d\varphi}$ i.e. local properties of the output state with respect to the parameter at a given parameter value $\varphi$. Saturating the C-R bound may therefore require
unrealistically good prior knowledge on the value of the estimated parameter. This is most pronounced by analyzing the behavior of
the phase estimation using the N00N states, which are invariant under $2\pi/N$ phase shifts and hence
require the prior knowledge of the parameter value to be of the order of $1/N$ as well.
Additionally, since $L_\varphi$ in general depends on $\varphi$ so can the optimal measurement, and again
a significant prior knowledge may be required to perform the optimal measurement. Last but not least, in order to quantify the performance
in terms of the \emph{total} resources consumed, i.e. $N_{\textrm{tot}}=kN$, one needs to know the behavior of the number of repetitions $k$ required to saturate the C-R bound with increase of $N_{\textrm{tot}}$, which is nontrivial and in general does not lead to analytical formulas. Specifically, to claim the Heisenberg limit in terms of $N_{\textrm{tot}}$, $k$ should not increase with $N_{\textrm{tot}}$ up to infinity \cite{Pezze2008}.

\section{Bayesian approach}
\label{sec:bayes}

An alternative analysis of the performance of quantum-enhanced measurement schemes, that does not suffer from the above mentioned deficiencies,
and hence yields the practically achievable precision limits, is the Bayesian approach where one explicitly takes into account the prior knowledge about the parameter value, represented by a probability distribution $p(\varphi)$ \cite{Buzek1999,Berry2000,Berry09,Kolodynski2010,Demkowicz2011}.
In this case, we define the average Bayesian error as
\begin{equation}\label{eq:bayes}
\overline{\Delta \varphi}=\sqrt{\int d\varphi \int dx p(\varphi)p_\Pi(x|\varphi)(\varphi - \tilde{\varphi}(x))^2}
\end{equation}
where $p_\Pi(x|\varphi)=\tr{\rho^N_\varphi\Pi_{x}}$. Here one averages error (precision) for some particular value of the parameter with the prior probability over the whole range of possible values of $\varphi$. In case of broad priors
one is therefore interested in finding strategies that work
globally, for many different values of the parameter, rather
than locally as in the case of the C-R bound based approach.
Finding the minimal $\overline{\Delta\varphi}$ requires optimization over input state, measurements and estimators which in general is much more demanding than maximization of the QFI over input states.
Yet, contrary to the QFI case, once the solution is found it yields a the explicit estimation procedure that saturates the minimal average Bayesian error.

Under certain regularity conditions one can relate the Bayesian and the C-R bound approaches through the so-called Bayesian C-R bound \cite{Gill1995}
\begin{equation}\label{eq:bcr}
\overline{\Delta \varphi}\geq  \frac{1}{ \sqrt{ \int \t{d}\varphi \, p(\varphi) F_\varphi  + \mathcal{I}}},
\end{equation}
where $\mathcal{I}=\int \t{d} \varphi \,\frac{1}{p(\varphi)} \left(\frac{\t{d} p(\varphi)}{\t{d} \varphi}\right)^2 $.
Provided the prior is smooth enough and it vanishes on the boundary of the set of allowed values of $\varphi$, the prior dependent term $\mathcal{I}$ is finite and
in the asymptotic limit of $N \rightarrow \infty$ becomes negligible as compared with $F_\varphi$. Moreover, in the unitary parameter estimation when the noise acts before the parameter encoding (or those two commute), i.e. when $\Lambda_\varphi^N(\cdot) = U_\varphi^{\otimes N} \Lambda^N (\cdot) U_\varphi^{\dagger \otimes n}$,  QFI does not depend on $\varphi$ \cite{Gaiba2009}, $F_\varphi = F$, irrespectively of the
presence or absence of decoherence, and hence \eref{eq:bcr} takes the form:
\begin{equation}
\overline{\Delta \varphi} \overset{N \rightarrow \infty}{\geq} \frac{1}{\sqrt{F}},
\end{equation}
implying that the Bayesian error is asymptotically also bounded by the standard C-R bound. We now ask whether it is possible to achieve equality in the above bound and hence prove asymptotic saturability of the C-R bound. Intuitively, this should be true because for very narrow priors $p(\varphi)$ one should be close to the local regime and both the C-R bound and the Bayesian approaches should give similar results. The same should keep in the asymptotic limit of large number of probes fed into the setup, because than the information gained from experiment is much larger than any a priori knowledge available in advance, the result known in classical parameter estimation as the Bernstein von-Misses theorem \cite{Vaart1998}.

\section{Estimation in the presence of decoherence}
\label{sec:decoherence}

\begin{figure}[t!]
\includegraphics[width=0.75\columnwidth]{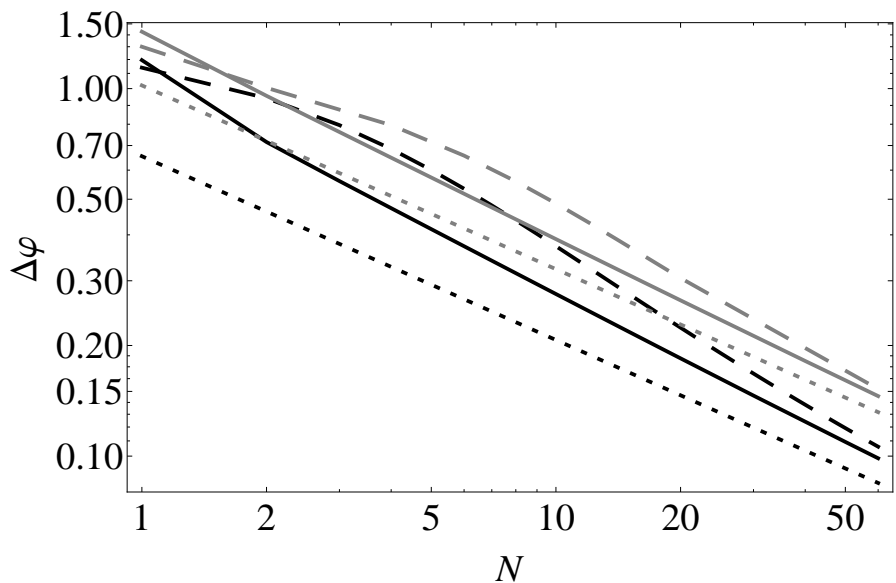}
\caption{Bayesian cost for the flat prior distribution $p(\varphi)=1/2\pi$ (dashed) vs. bound given by the QFI (solid) as a function of the number of particles for losses (black) and local dephasing (gray) with decoherence parameter $\eta=0.7$. For comparison ultimate asymptotic QFI based bounds on precision \cite{Demkowicz2012,Escher2011,Knysh2014} are depicted for losses $\sqrt{\frac{1-\eta}{\eta N}}$ (black, dotted) and dephasing $\sqrt{\frac{1-\eta^2}{\eta^2 N}}$ (gray, dotted).}
\label{fig:bayesqfi}
\end{figure}

Let us consider first the situation when the maximal QFI scales asymptotically at most linearly with $N$,
$F \overset{N \rightarrow \infty}{=}  \alpha  N$, which is a generic case for metrological
models with uncorrelated noise \cite{Matsumoto2010, Fujiwara2008, Demkowicz2012}.
 Since QFI is additive on product states, $F(\rho^{\otimes k})=kF(\rho)$,  it implies that for a sufficiently large $N$ instead of taking a general entangled state of $N$ particles $\ket{\psi_N}$, one could take separable state of
$k$ copies ("groups") of an entangled state $\ket{\psi_n}$ with smaller number of particles $n=N/k$ and achieve almost the same QFI.
More formally, let us expand the optimal asymptotic QFI in powers of $N$ taking into account the leading correction to the
linear asymptotic scaling which without loss
of generality may be written as $F(N)\approx N(\alpha - \beta N^{-\gamma})$, see e.g. \cite{Knysh2011, Knysh2014}, with $\beta, \gamma >0$.
The grouping procedure would not change the optimal QFI by more than $\epsilon$, $k F(n)/F(N) \geq 1-\epsilon$, provided the size of the group
satisfies:
\begin{equation}
n \geq \left(\frac{\alpha \epsilon}{\beta} +(1-\epsilon) N^{-\gamma}\right)^{-1/\gamma} \overset{N \rightarrow \infty}{=}
\left(\frac{\beta}{\alpha\epsilon}\right)^{1/\gamma},
\end{equation}
which implies that for any $\epsilon>0$ the size of the group $n$ can be assumed to be finite in the asymptotic limit $N \rightarrow \infty$,
while the number of groups $k$ grows to infinity proportionally to $N$.
Therefore the estimation problem in the asymptotic limit, can be effectively viewed as a parameter estimation problem
on a large number of independent and identical copies---$(\rho^n_\varphi)^{\otimes k}$. In this case, however,
under some regularity conditions for the Bayesian model \cite{Gill2005, Gill2013} there exist a Bayesian estimation strategy that
is asymptotically efficient and saturates the C-R bound. For this purpose one can e.g. refer to an elegant
quantum local asymptotic normality theorem \cite{Guta2007, Kahn2009} which states  that in the asymptotic limit
the estimation problem on uncorrelated copies may be equivalently viewed as an estimation problem on a multi mode quantum Gaussian states
with the estimated parameter being encoded in a displacement of the state.  The optimal estimation strategy then amounts to a measurement
of a particular quadrature operator yielding Gaussian probability distribution with the variance determined by the QFI.
This proves that the QFI based bound \eref{eq:crdecoh} is indeed asymptotically saturable and allow us to rewrite it
 as an equality for the asymptotically achievable Bayesian cost with the constant in the enumerator unchanged:
\begin{equation}
\overline{\Delta \varphi} \overset{N \rightarrow \infty}{=} \frac{\t{const}}{\sqrt{N}}.
\end{equation}
As an example, in Fig.~(\ref{fig:bayesqfi}), we depict the precision limits for
phase estimation on $N$ 2-level systems under two different decoherence models: (i) losses or (ii)
uncorrelated dephasing,
where it is clearly seen that for large $N$ respective Bayesian cost and bound given by the QFI indeed converge.
Discussion of the effective numerical approach that allows to obtain exact results for large number of particles
and the details of the models are discussed in the appendices.

We have mentioned before that linear scaling of QFI, which is a crucial assumption in the above equivalence argument, is 
generic in models with uncorrelated noise. However, in problems where the decoherence strength may be tuned 
with the increase of $N$, as e.g. in frequency estimation schemes where one is allowed to optimize over the probes interrogation time, the situation may be different. This is the case in e.g. perpendicular dephasing \cite{Chaves2013} or non-Markovian evolution \cite{Chin2012} models, where in the limit of increasing number of probes, a choice of properly decreasing interrogation times may effectively reduce the impact of decoherence and allow the QFI to scale better than linearly. In such cases a dedicated analysis, which is beyond the scope of the present paper, is required in order to relate the Bayesian and the C-R bound approaches.

\section{Decoherence-free estimation}
\label{sec:nodecoh}

Let us now consider the decoherence-free case, $\Lambda_\varphi=U_\varphi$. Since QFI scales quadratically with $N$ we can
no longer apply the previous argument about asymptotic "group" structure of the optimal input state.
Interestingly, for phase estimation, $U_\varphi = e^{\mathrm{i} \sigma_z \varphi/2}$, and the flat prior, $p(\varphi)=1/2\pi$,
 it is possible to derive analytically the optimal Bayesian solution utilizing the concept of covariant measurements, see \ref{app:a}, which
 asymptotically yields $\overline{\Delta \varphi} \overset{N \rightarrow \infty}{=} \pi/N$ \cite{Berry2000}.
 This asymptotic result is by a factor of $\pi$ larger from value of the respective C-R bound, see \eref{eq:crunitary}.
 One might argue that this discrepancy arises due to the assumption of the flat prior in the Bayesian approach and that by narrowing the prior
 one might eventually achieve the exact $1/N$ scaling. We show below, that this intuition is wrong, by considering arbitrarily narrow Gaussian
 priors and proving that the asymptotic scaling remains $\pi/N$, which demonstrates that the C-R bound is not achievable in this case.

\begin{figure}[t!]
\includegraphics[width=0.75\columnwidth]{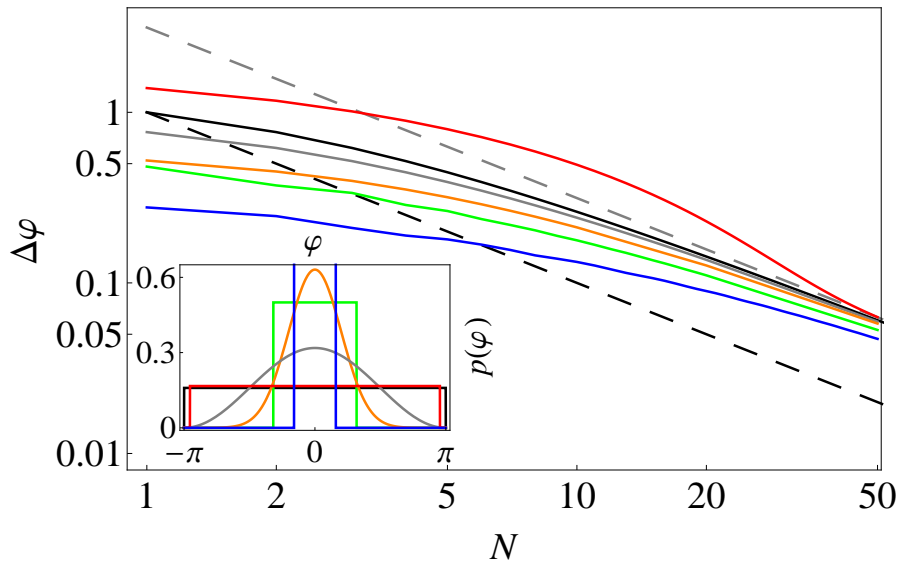}
\caption{Bayesian cost for decoherence-free phase estimation for various prior distribution $p(\varphi)$ all asymptotically converge to $\pi/N$ formula (gray, dashed).  
For comparison, $1/N$ C-R bound is given by black dashed line. The shapes of the prior distribution are depicted in the inset.}
\label{fig:global}
\end{figure}

Consider a Gaussian prior $p(\varphi)=\frac{1}{\sqrt{2\pi\Delta_0^2}}e^{-\varphi^2/2 \Delta_0^2}$ and assume that the width of the prior
distribution $\Delta_0 \ll 1$, so that
 it is narrow enough so we can neglect the tails outside the interval $(-\pi,\pi)$.
For unitary parameter estimation with Gaussian prior and quadratic cost there is a close relation between the
Bayesian cost and the QFI \cite{Macieszczak2013}:
\begin{equation}
\label{eq:bayesfisher}
\overline{\Delta\varphi}=\Delta_0 \sqrt{1-\Delta^2_0 F(\bar{\rho})},
\end{equation}
where $F(\bar{\rho})$ is the QFI
calculated for the prior-averaged probe state $\bar{\rho}=\int \t{d} \theta \,p(\theta)U_\theta^{\otimes N}\ket{\psi_N}\bra{\psi_N} U_\theta^{\dagger \otimes N} $. Looking for the minimal $\overline{\Delta \varphi}$ is therefore equivalent to determining the input state
for which  $F(\bar{\rho})$ is maximal.
Since $\bar{\rho}$ may also be formally viewed as the input probe state subjected to collective dephasing
we can utilize the asymptotic formula for the optimal QFI for phase interferometry under collective dephasing derived in \cite{Knysh2014} which reads
 $F=\frac{1}{\Gamma+\pi^2/N^2}$, where  the dephasing strength parameter $\Gamma$  needs to be replaced with the prior variance $\Delta_0^2$. Substituting this result into \eref{eq:bayesfisher} we get
\begin{equation}
\overline{\Delta \varphi} =\sqrt{\Delta_0^2\left(1-\frac{\Delta_0^2}{\Delta_0^2+\pi^2/N^2}\right)} \overset{N \rightarrow \infty}{=} \frac{\pi}{N}
\end{equation}
irrespectively of the width of the prior distribution. The assumption of Gaussianity of the prior was needed for formal derivation of the
above result but we conjecture that the above holds for general sufficiently regular prior distributions.
This is intuitively obvious since we get the same results for flat prior and all Gaussian priors including arbitrary narrow ones.
Therefore it is natural to expect that all intermediate cases should manifest the same behavior. This means that in the decoherence-free case correct limit on the phase estimation error is given by $\pi/N$, and not $1/N$. Numerically results confirming this reasoning, obtained using the techniques of \cite{Demkowicz2011}, are illustrated Fig.~\ref{fig:global}.

 Moreover, based on numerical calculations, we conjecture that
 the $\pi$ factor discrepancy between the C-R bound and
 the asymptotically saturable precision derived for the phase estimation problem holds in general for any decoherence-free unitary parameter estimation $U_\varphi=e^{-i\varphi H}$, and the correct form of the optimal asymptotically achievable uncertainty reads:
\begin{equation}
\label{eq:unitary}
\overline{\Delta \varphi} \overset{N \rightarrow \infty}{=} \frac{\pi}{(\lambda_+-\lambda_-)N}
\end{equation}
irrespectively of the prior. Intuitively, by performing preliminary measurements on negligible portion of the particles we can narrow
the prior distribution to have width of the order of $2\pi /(\lambda_+ - \lambda_-)$ so that we will not suffer estimation ambiguity
due to using eigenstates with just the extremal eigenvalues. After this preliminary procedure the optimal
strategy is isomorphic to the Bayesian phase estimation strategy up to the rescaling of the phase evolution speed by $\lambda_+ - \lambda_-$.
Formula \eref{eq:unitary} should thus be regarded as a refinement of the previously derived C-R based bounds for unitary
parameter estimation \cite{Giovannetti2006}

\section{Estimation in the presence of global dephasing}
\label{sec:global}

In the above discussion we have considered models where channels $\Lambda_\varphi$ acting on different particles where uncorrelated, as in Fig.~\ref{fig:scheme}b.
There are situations, however, when the setup cannot be decomposed into separate channels acting on each of the probes. This may be caused by the presence of the memory or some long distance correlations between channels. In such case making any general statement about asymptotic value of precision is non-trivial. In fact, as we will show, there are some cases in which one cannot define asymptotic value of precision without paying enough attention to the form of a priori probability distribution and hence cannot make a meaningful connection between Bayesian and C-R bound based approaches.

\begin{figure}[t!]
\includegraphics[width=0.75\columnwidth]{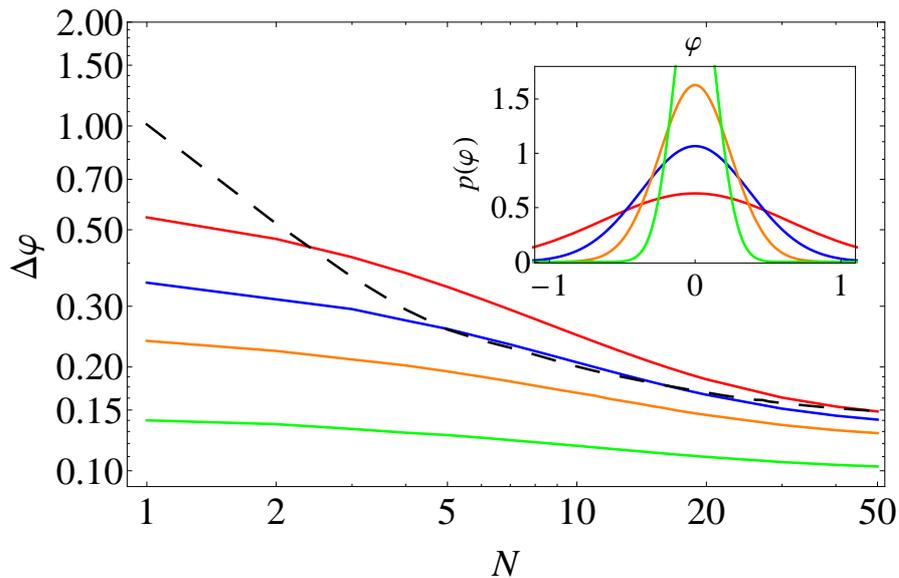}
\caption{Bayesian cost for various prior distribution $p(\varphi)$ (solid) as a function of the number of particles for collective dephasing with decoherence strength $\Gamma=0.02$. Asymptotic precision clearly depends on the prior and in general does not coincide with the asymptotic value
of the value of the QFI (dashed).}
\label{fig:collective}
\end{figure}

As an illustrative example, consider a phase estimation problem in presence of collective dephasing,
so that
\begin{equation}
\rho_\varphi^N =  U_\varphi^{\otimes N} \left( \int \t{d} \theta  q(\theta) U_\theta^{\otimes N} \ket{\psi_N}\bra{\psi_N}  U_\theta^{\dagger \otimes N }\right) U_\varphi^{\otimes N \dagger}
\end{equation}
where $U_\varphi= e^{\mathrm{i} \varphi \sigma_z/2}$ and $q(\theta)=\frac{1}{\sqrt{2\pi\Gamma}} e^{-\theta^2/2\Gamma}$ with $\Gamma$ being the dephasing strength parameter.
 For Gaussian prior $p(\varphi)= \frac{1}{\sqrt{2\pi\Delta_0^2}} e^{-\varphi^2/2\Delta_0^2}$ we may again utilize equation \eref{eq:bayesfisher} but this time while calculating QFI, $F(\bar{\rho})$, the averaged state $\bar{\rho}$ needs to be effectively phase averaged both due to prior distribution as well as  the actual dephasing process. For large $N$ this yields $F(\bar{\rho}) = \frac{1}{\Gamma +\Delta_0^2 + \pi^2/N^2}$ as
 the convolution of two Gaussian distributions is again a Gaussian with a variance being the sum of the two. Plugging this formula into
 \eref{eq:bayesfisher} we find the formula for the optimal Bayesian cost.
 \begin{equation}
 \overline{\Delta \varphi} =\Delta_0 \sqrt{1-\frac{\Delta_0^2}{\Gamma + \Delta_0^2+\pi^2/N^2}} \overset{N \rightarrow \infty}{=}
 \sqrt{\frac{\Gamma}{1+\Gamma/\Delta_0^2}},
 \end{equation}
 showing a clear dependence on the prior knowledge except in the case when $\Gamma \ll \Delta_0^2$ in which case the C-R based and the Bayesian
 approaches predict the same asymptotic value for precision equal $\sqrt{\Gamma}$.
 This is due to the fact that the information on the estimated parameter does not increase indefinitely with $N$ and hence in the asymptotic limit the estimator will not approach the true value of the parameter. Then, it should be no surprise that for peaked prior distribution $\Delta_0^2\ll\Gamma$ the prior will dominate the  resulting optimal precision, and hence no asymptotic prior independent formula for precision exists, see Fig.~\ref{fig:collective}. The same behavior will also be observed if collective dephasing is added on top of uncorrelated decoherence processes.

Above reasoning clearly shows that the bound based on the QFI
is of limited use in a ``single-shot'' analysis of setups under collective decoherence
and a more suitable measure of precision is the Bayesian cost. It should be noted, however,
that in practice one would avoid performing single shot
experiments employing states with large number of particles $N$
subject to collective decoherence. Instead, a preferable strategy would be to divide $N$ 
into $k$ groups with smaller number of particles $n=N/k$ and send
them separately, making the collective decoherence act on
each of the groups individually and thus restoring precision
$\Delta\varphi\sim \frac{\Delta\varphi_{(n)}}{\sqrt{k}}$, where
$\Delta\varphi_{(n)}$ is the precision obtained with the one group \cite{Knysh2014}. 
In this case, arguments presented in Sec.~\ref{sec:decoherence} when discussing
problems with uncorrelated noise apply, and the Bayesian prior
will lose its significance in the limit of many experiment
repetitions $k \rightarrow \infty$.


\section{States with indefinite number of particles}
\label{sec:indefinite}
States with indefinite number of particles, such as coherent or squeezed states, are a natural candidates for
optical implementations of quantum metrological schemes as they are relatively easy to prepare with the state-of-the-art technology.
In particular interference of squeezed and coherent states is at the moment the only feasible technique allowing to benefit
from the quantum features of light in devices operating in the large light intensity regime such as gravitational wave detectors \cite{LIGO2013}. Interestingly, such protocols despite their conceptual simplicity offer practically optimal performance from the point of view of quantum-enhancement effects \cite{Demkowicz2013}. Still, mathematical analysis of ultimate performance of protocols utilizing
states with indefinite number of particles is more involved than for states with definite particle number.
Considering such states one has first to decide whether quantum coherences between sectors of Hilbert space representing
different total photon numbers are observable. We here take the position that
observability of such coherences necessarily require presence of an additional phase reference beam, which should therefore be counted in
as resources in any interferometric experiment \cite{Bartlett2007, Jarzyna2012, Demkowicz2014}. If the reference beam is not explicitly included in the resources it should be regarded as absent and the state with an average photon number $\bar{N}$
should be effectively treated as being an incoherence mixture (a direct sum in this case)
of different definite-photon number states:
\begin{equation}
\rho^{\bar{N}} =\bigoplus_{N=0}^{\infty} p_N \rho^N, \quad \sum_{N} p_N N = \bar{N}.
\end{equation}
Note, that we may also consider such states in case of protocols involving massive particles, for which coherent superposition
of different particle number states is forbidden by the superselection rules, as they may represent
a probabilistic scheme with different definite particle number state prepared with different probabilities.
Importantly, most of the discussions that arise around the utility of states with indefinite particle number and in particular the feasibility
of sub-Heisenberg strategies can be restricted to this class of states as the essence of the problem lies in possibility of
mixing and not superposing different particle number states.

The simplest example of reasoning based on the use of the QFI that can lead to claims on sub-Heisenberg precision in phase estimation involves
a state which is mixture of a vacuum state and the $N$ photon N00N state. As the terms in the mixture occupy orthogonal subspaces,
the QFI for such a mixture is a weighted sum of QFIs for each of the constituents $F =(1- p) 0 + p N^2$ where $p$ is the probability of sending the N00N state. The average photon number $\bar{N} = pN$ is treated as a fixed resource, and we can rewrite the QFI in the form $F=  \bar{N} N$.
Hence for fixed $\bar{N}$ we may increase $N$ indefinitely making QFI arbitrary large, which translates to C-R bound with arbitrary small
estimation uncertainty. In practice, however, these type of strategies require prior knowledge that also increases with $N$ making
them not practical \cite{Tsang2012, Giovanetti2012}. Building on techniques presented in this paper, we show below an alternative argument
that states with indefinite particle numbers indeed offer no asymptotic benefits over definite particle number strategies.

For optimal Bayesian phase estimation with a Gaussian prior we may again utilize formula \eref{eq:bayesfisher} as
in its derivation no assumption on the particle-definiteness was ever assumed. The optimal Bayesian cost for
$\rho^{\bar{N}}$ state thus read:
\begin{equation}
\Delta\varphi_{\bar{N}} = \Delta_0 \sqrt{1 - \Delta_0^2 F(\bar{\rho}^{\bar{N}})}.
\end{equation}
Since $\bar{\rho}^{\bar{N}}$ just as $\rho^{\bar{N}}$ is a mixture of states occupying orthogonal subspaces, we can write:
$F(\bar{\rho}^{\bar{N}}) = \sum_N p_N F(\bar{\rho}^N)$. As we will be interested in large $\bar{N}$ regime, let us first assume
that all the relevant terms in the mixture correspond to large $N$ so that the large $N$ approximation to
the optimal QFI of the $N$ particle dephased state $\max_{\bar{\rho}^N} F(\bar{\rho}^N) = \frac{1}{\Delta^2_0 + \pi^2/N^2}$ hold.
Hence we can write
\begin{equation}
\label{eq:indefinite}
\Delta\varphi_{\bar{N}} \geq  \Delta_0 \sqrt{1 - \Delta_0^2 \sum_N  \frac{p_N}{\Delta^2_0 + \pi^2/N^2}} \geq
\Delta_0 \sqrt{1 - \frac{\Delta_0^2 }{\Delta^2_0 + \pi^2/\bar{N}^2}}  \overset{\bar{N} \rightarrow \infty}{=} \frac{\pi}{\bar{N}},
\end{equation}
where in the last inequality we have made use of the concavity property of $1/(1+1/x^2)$ function.
This clearly demonstrates that there is no benefit in using mixtures as the cost will only be higher than the cost corresponding
to a definite particle state with $N=\bar{N}$. A missing point in the above reasoning is the assumption that all relevant constituents of the mixture correspond to large $N$. This was clearly not the case in the elementary example presented before involving a mixture of the vacuum state.
Assume then that there is a finite $M$ such that for states with particle numbers $N<M$ have a finite weight $p>\epsilon$.
Then the cost $\Delta\varphi_{\bar{N}} \geq \epsilon \Delta\varphi_{M}$, as the optimal cost for each of the
$N<M$ terms cannot be smaller than for an optimal $M$ particle state whereas including states with $N>M$ only
increases the cost. Therefore, when increasing $\bar{N}$ and heading for better precision we must necessarily decrease $\epsilon$
or increase $M$ in order not to be bounded from below be a finite uncertainty. This implies that the only way to have estimation uncertainty that asymptotically goes to zero, is to only deal with mixtures where effectively all weight is carried by states with increasing $N$, and asymptotically no finite weight may be kept in below any finite $M$. This supports the reasoning leading to \eref{eq:indefinite} and excludes the possibility of
better than Heisenberg scaling of precision within the Bayesian approach.

\section{Conclusions}
\label{sec:conclusions}

In summary, we have proven that in the presence of uncorrelated decoherence the asymptotic limits on precision
of quantum metrological schemes may be credibly calculated using the C-R bound based approach whereas
in the decoherence-free unitary parameter estimation a $\pi$ factor correction
needs to be included irrespectively of the extent of prior knowledge.
These observations provide a firm ground for the use of the QFI as a sensible figure of merit in analyzing the performance of quantum enhanced metrological protocols based on definite-particle number states. In case of strategies employing states with indefinite number of particles
the claims remain unchanged in presence of uncorrelated noise. In the decoherence-free case, however, the Bayesian analysis
shows that C-R bound motivated proposals of sub-Heisenberg estimation strategies
are not of much practical use, and the actual Bayesian cost cannot scale better than $\pi/\bar{N}$ where $\bar{N}$ is the average number of particles.

\section*{Acknowledgements}

We would like to thank Janek Ko{\l}dy{\'n}ski, Madalin {Gu{\c t}{\u a}} and Lorenzo Maccone for helpful comments and fruitful discussions. This research was supported by the EC under the FP7 IP project SIQS co-financed by the Polish Ministry of Science and Higher Education.

\appendix

\section{Optimization of the Quantum Fisher Information and the Bayesian cost}
\label{app:a}
Here we  briefly  discuss methods which allowed us to compute the Quantum Fisher Information and the Bayesian cost efficiently for large number of probes in the presence of decoherence. In the first case we have used an iterative algorithm proposed by \cite{Macieszczak2013,Macieszczak2013a} which may be summarized as follows:


\begin{enumerate}
\item
Take some reasonable initial state $\ket{\psi^{(0)}}$ and calculate for it the output density matrix $\rho^{(0)}$ and SLD $L^{(0)}$.
\item
Calculate operator $A=\Lambda^*(L^{(0)2}-2i[H,L^{(0)}])$, where $H$ is the generator of the unitary evolution which encodes the parameter and $\Lambda^*$ represents the channel in the Heisenberg picture $\Lambda^*(A)=\sum_i K_i^{\dagger}AK_i$.
\item
Find the eigenvector $\ket{v}$ of $A$ corresponding to the smallest eigenvalue.
\item
Take $\ket{\psi^{(1)}}=\ket{v}$ and repeat the procedure.
\end{enumerate}
After sufficiently large amount of iterations, such procedure would give almost optimal state, for which one can calculate the Quantum Fisher Information.

In the case of Bayesian cost in calculations we have used slightly more general form of the error, i.e.
\begin{equation}
\overline{\Delta \varphi}=\sqrt{\int d\varphi \int dx p(\varphi)p_\Pi(x|\varphi)c(\tilde{\varphi}(x),\varphi)}
\end{equation}
where $c(\tilde{\varphi}(x),\varphi)$ is called a cost function. Here we considered two types of cost functions: quadratic cost function $c(\tilde{\varphi},\varphi)=\left(\tilde{\varphi}-\varphi\right)^2$ and sine cost function $c_s(\tilde{\varphi},\varphi)=4\sin\left(\frac{\tilde{\varphi}-\varphi}{2}\right)^2$, the latter one naturally emerging for the problem of phase estimation due to periodicity of the parameter (note that $c_s(\tilde{\varphi},\varphi)\approx c(\tilde{\varphi},\varphi)$ whenever $\tilde{\varphi}\approx\varphi$ so asymptotically Bayesian cost for sine cost function should be equal to that calculated with $c(\tilde{\varphi},\varphi)$). For the problems we have considered, dealing with the first function is hard and in general possible only numerically. On the other hand, the sine cost function greatly simplifies the problem for phase estimation and flat a priori knowledge since one can restrict measurements to a class of covariant POVMs \cite{Holevo1982,Chiribella2005,Bartlett2007} parametrized by the estimated value and given by
\begin{equation}
\Pi_{\tilde{\varphi}}=U_{\tilde{\varphi}}\Xi U_{\tilde{\varphi}}^{\dagger},\quad \int_{-\pi}^{\pi}\frac{d\tilde{\varphi}}{2\pi}U_{\tilde{\varphi}}\Xi U_{\tilde{\varphi}}^{\dagger}=1\!\!1,
\end{equation}
where $\Xi$ is a positive semi-definite operator called the seed operator. Using the above formula, the average cost simplifies to
\begin{equation}\label{eq:BayesCov}
\overline{\Delta\varphi}=4\int_{-\pi}^{\pi}\frac{d\varphi}{2\pi}\tr{\rho_\varphi\Xi}\sin^2\frac{\varphi}{2}.
\end{equation}
With the help of the above equation one may easily derive average cost for the decoherence-free estimation \cite{Berry2000}, losses \cite{Kolodynski2010}, global dephasing \cite{Demkowicz2014} or local dephasing (see \ref{app:c}).

In general, for other types of prior probability distributions and more general unitary transformations one have to use iterative algorithms similar to the one described above which are described in details either in \cite{Demkowicz2011} for the sine cost function and in \cite{Macieszczak2013} for the quadratic cost function.


\section{Derivation of the density matrix in the presence of depahsing}
\label{app:b}
Iterative algorithms described above reduce the optimization problem to a repeated solving of a matrix eigenproblem. Still, in order to fully utilize them one needs to efficiently describe the output density matrices. This is particularly challenging in the case of local dephasing where the output density matrix lies outside the fully symmetric subspace and its dimension in principle scales exponentially with the number of probes.
Here we derive a way to efficiently describe the output density matrix for interferometric models under dephasing or loss for $N$ two-level input probes prepared initially in a symmetric state, which can in general be  written in the bosonic mode occupation notation $\ket{\psi_N}=\sum_{n=0}^{N}c_n\ket{n,N-n}$.

Local dephasig can be described using two single-particle Kraus operators of the form
\begin{equation}
K_0=\sqrt{\frac{1+\eta}{2}}1\!\!1 
,\quad
K_1=\sqrt{\frac{1-\eta}{2}}\sigma_z,
\end{equation}
where $\eta$ denotes strength of decoherence and $\sigma_z$ is the Pauli $z$ operator. The $N$ particle output density matrix is equal to
\begin{eqnarray}
\fl
\nonumber
\rho^N = \Lambda^N[\ket{\psi_N}]=\sum_{k=0}^{N}\sum_{\pi_k^N}\pi_k^N\left(K_1^{\otimes k}\otimes K_0^{\otimes N-k}\right)\\
\ket{\psi_N}\bra{\psi_N}\pi_k^N\left(K_1^{\dagger\otimes k}\otimes K_0^{\dagger\otimes N-k}\right),
\end{eqnarray}
where $\pi_k^N$ represents different permutations of $k$ and $N-k$ copies of $K_1$ and $K_0$ operators respectively. To simplify the problem, we can treat our two-level probes as spin $1/2$ particles and set up a notation in which we write the input state as a state with total angular momentum $j=\frac{N}{2}$. Then, its $z$ components are equal to $m=n-\frac{N}{2}$, $\ket{n,N-n}=\ket{\frac{N}{2},m}$. We may now utilize the well
known techniques for adding angular momenta and rewrite:
\begin{equation}
\ket{\frac{N}{2},m}=\sum_{k=0}^{N}c_{\frac{k}{2},\tilde{m},\frac{N}{2},m-\tilde{m}}^{\frac{N}{2},m}\ket{\frac{k}{2},\tilde{m}}\ket{\frac{N-k}{2},m-\tilde{m}}
\end{equation}
where $c_{\frac{k}{2},\tilde{m},\frac{N}{2},m-\tilde{m}}^{\frac{N}{2},m}=\bra{\frac{k}{2},\tilde{m}}\bra{\frac{N-k}{2},m-\tilde{m}}|\frac{N}{2},m\rangle$ are the Clebsch-Gordan coefficients. Thus
\begin{eqnarray}
\nonumber
K_1^{\otimes k}\otimes K_0^{\otimes N-k}\ket{\frac{N}{2},m}=\\
\nonumber\sum_{k=0}^{N}c_{\frac{k}{2},\tilde{m},\frac{N-k}{2},m-\tilde{m}}^{\frac{N}{2},m}\left(\frac{1-\eta}{2}\right)^{\frac{k}{2}}\left(\frac{1+\eta}{2}\right)^{\frac{N-k}{2}}\cdot\\
 \nonumber \cdot (-1)^{\frac{k}{2}-\tilde{m}}\ket{\frac{k}{2},\tilde{m}}\ket{\frac{N-k}{2},m-\tilde{m}}=\\ \nonumber
=\sum_{k=0}^{N}\sum_{j=|k-\frac{N}{2}|}^{\frac{N}{2}}\sum_{\alpha_j}c_{\frac{k}{2},\tilde{m},\frac{N}{2},m-\tilde{m}}^{\frac{N}{2},m}c_{\frac{k}{2},\tilde{m},\frac{N}{2},m-\tilde{m}}^{j,m}\cdot \\
 \cdot\left(\frac{1-\eta}{2}\right)^{\frac{k}{2}}\left(\frac{1+\eta}{2}\right)^{\frac{N-k}{2}}(-1)^{\frac{k}{2}-\tilde{m}}\ket{j,m,\alpha_j},
\end{eqnarray}
where $\alpha_j$ denotes multiplicity of the subspace with total angular momentum $j$. Eventually, we may express the output density matrix as
\begin{eqnarray}
\nonumber
\rho^N=\sum_{m,m'}\rho_{m,m'}\Lambda^N\left(\ket{\frac{N}{2},m}\bra{\frac{N}{2},m'}\right)\to\\ \nonumber
\to\sum_{k=0}^N\sum_{j,j'=|\frac{N}{2}-k|}^{N/2}\sum_{m,m'=-j,-j'}^{j,j'}\sum_{\alpha_j,\alpha_{j'}}\left(\frac{1-\eta}{2}\right)^{k}\left(\frac{1+\eta}{2}\right)^{N-k}\\ \nonumber
\rho_{m,m'}C_{j,m}^{N,k}C_{j,m'}^{N,k}\sum_{\Pi_k^N}\Pi_k^N\left(\ket{j,m,\alpha_j}\bra{j',m',\alpha_{j'}}\right)=\\ \nonumber
=\sum_{j,j'=0}^{N/2}\sum_{m,m'=-j}^{j}\sum_{k=\frac{N}{2}-j}^{\frac{N}{2}+j}{N \choose k}\left(\frac{1-\eta}{2}\right)^{k}\left(\frac{1+\eta}{2}\right)^{N-k}\\
\rho_{m,m'}C_{j,m}^{N,k}C_{j,m'}^{N,k}\ket{j,m}\bra{j,m'}\otimes \frac{1}{d_j}1\!\!1_{\mathbb{C}_{d_j}}
\end{eqnarray}
where
\[
C_{j,m}^{N,k}=\sum_{\tilde{m}=-k/2}^{k/2}c_{\frac{k}{2},\tilde{m},\frac{N}{2},m-\tilde{m}}^{\frac{N}{2},m}c_{\frac{k}{2},\tilde{m},\frac{N}{2},m-\tilde{m}}^{j,m}(-1)^{\frac{k}{2}-\tilde{m}}
\]
and $d_j$ is the dimension of the multiplicity space corresponding to the total angular momentum $j$.
Since the multiplicity subspaces are not affected by the phase-sensing transformation $U_\varphi^{\otimes N}$
we can ignore them and effectively write $\rho^N$ in a block diagonal form:
\begin{equation}\label{eq:StateDeph}
\fl
\rho^N = \bigoplus_{j=0}^{N/2}\sum_{m,m'=-j}^{j}\sum_{k=\frac{N}{2}-j}^{\frac{N}{2}+j}{N \choose k}\left(\frac{1-\eta}{2}\right)^{k}\left(\frac{1+\eta}{2}\right)^{N-k}\rho_{m,m'}C_{j,m}^{N,k}C_{j,m'}^{N,k}\ket{j,m}\bra{j,m'}
\end{equation}
Equation \eref{eq:StateDeph} describes density matrix with dimension equal to $\left(\frac{N}{2}+1\right)^2$ and scales only quadratically in the number of probes compared to exponential scaling for "brute force" description. Similar formula but utilizing spherical tensors was also found in \cite{Frowis2014}.

The case of losses is relatively simpler. We model loss of probes by inserting two artificial beam splitters in both arms of the interferometer with transmissivities $\eta$ and vacuum states fed into the respective second input ports. By a standard beam splitter transformation and tracing out the environment one may easily derive the output density matrix as \cite{Demkowicz2009}
\begin{equation}
\rho^N=\Lambda^N[\ket{\psi_N}] = \sum_{l_0=0}^N\sum_{l_1=0}^{N-l_0}p_{l_0l_1}\ket{\psi_{l_0l_1}}\bra{\psi_{l_0l_1}}
\end{equation}
where
\begin{eqnarray}
\ket{\psi_{l_0l_1}}=\frac{1}{\sqrt{p_{l_0l_1}}}\sum_{n=l_0}^{N-l_1}c_n B_{l_0l_1}^n(\eta)\ket{n-l_0,N-n-l_1}\\
B_{l_0l_1}^n(\eta)=\sqrt{{n \choose l_0}{N-n\choose l_1}\eta^{N-l_0-l_1}(1-\eta)^{l_0+l_1}}
\end{eqnarray}
and $p_{l_0l_1}$ is a normalization factor and $l_0,\,l_1$ represents number of photons lost in respective arms. Such a density matrix has dimension $(N+1)(N+2)/2$ which is again quadratic in the number of probes and thus is feasible to use in iterative procedures.

\section{Bayesian cost in the presence of local dephasing}
\label{app:c}

Using formulas \eref{eq:BayesCov} and \eref{eq:StateDeph} we may derive the Bayesian cost for the flat prior and the sine cost function in the presence of local dephasing. Calculations are similar to the case of losses obtained in \cite{Kolodynski2010}. Because the output density matrix is block-diagonal $\rho=\oplus_{j=0}^{N/2} \rho^j$,  without loss of generality we may assume that our seed operator is also block diagonal $\Xi=\oplus_{j=0}^{N/2} \Xi^j$. Equation \eref{eq:BayesCov} can now be written as:
\begin{eqnarray}\nonumber
\overline{\Delta\varphi}=4\sum_{j=0}^{N/2}\int_{-\pi}^{\pi}\frac{d\varphi}{2\pi}\tr{U_\varphi\rho^j U_\varphi^{\dagger}\Xi^j}\sin^2\frac{\varphi}{2}=\\ \nonumber
=4\sum_{j=0}^{N/2}\int_{-\pi}^{\pi}\frac{d\varphi}{2\pi}\sum_{m,m'=-j}^j\rho^j_{m,m'}\Xi^j_{m',m}e^{-i\varphi(m-m')}\sin^2\frac{\varphi}{2}=\\
=\sum_{j=0}^{N/2}\sum_{m,m'=-j}^j\rho^j_{m,m'}\Xi^j_{m',m}f_{m-m'}
\end{eqnarray}
where $f_{n}=4\int_{-\pi}^{\pi}\frac{d\varphi}{2\pi}e^{-i\varphi n}\sin^2\frac{\varphi}{2}$. Note that the only nonzero elements are $f_0=2$ and $f_{\pm 1}=-1$. Now, substituting \eref{eq:StateDeph} into the above equation, gives us
\begin{eqnarray}
\fl
\nonumber
\overline{\Delta\varphi}=2+\sum_{j=0}^{N/2}\sum_{m\neq m'=-j}^j c_m c_{m'}^{*}A^{N,j}_{m,m'}(\eta)\Xi^j_{m',m}f_{m-m'}\\ \nonumber
\geq 2+\sum_{j=0}^{N/2}\sum_{m\neq m'=-j}^j |c_m||c_{m'}|A^{N,j}_{m,m'}(\eta)|\Xi^j_{m',m}|f_{m-m'}\\
\geq 2+\sum_{j=0}^{N/2}\sum_{m\neq m'=-j}^j |c_m||c_{m'}|A^{N,j}_{m,m'}(\eta)f_{m-m'}
\end{eqnarray}
where we have used shorthand notation $A^{N,j}_{m,m'}(\eta)=\sum_{k=\frac{N}{2}-j}^{\frac{N}{2}+j}{N \choose k}\left(\frac{1-\eta}{2}\right)^{k}\left(\frac{1+\eta}{2}\right)^{N-k}C_{j,m}^{N,k}C_{j,m'}^{N,k}$. The first inequality comes from the fact that $f_{\pm 1}<0$ and the second one form $\Xi^j_{m',m}\leq\sqrt{\Xi^j_{m,m}\Xi^j_{m'm'}}$ which is a consequence of positive semidefiniteness of the seed operator. Both of these inequalities are saturated by $\Xi^j=\ket{e_j}\bra{e_j}$ where $\ket{e_j}=\sum_{m=-j}^j\ket{j,m}$.

Now, using the optimal seed operator $\Xi=\oplus_{j=0}^{N/2}\ket{e_j}\bra{e_j}$ we may write that
\begin{equation}
\overline{\Delta\varphi}=2-\mathbf{c}^T M\mathbf{c},
\end{equation}
where $\mathbf{c}$ represents the vector of state coefficients and $M$ is matrix with nonzero entries
\begin{equation}
M_{m,m+1}=M_{m+1,m}=\sum_{j=\frac{N}{2}+m+1}^{N/2}A^{N,j}_{m,m+1}(\eta)
\end{equation}
Finding Bayesian cost reduces therefore to finding the largest eigenvalue of the matrix $M$.

\bibliographystyle{iopart-num}

\begin{thebibliography}{10}
\expandafter\ifx\csname url\endcsname\relax
  \def\url#1{{\tt #1}}\fi
\expandafter\ifx\csname urlprefix\endcsname\relax\def\urlprefix{URL }\fi
\providecommand{\eprint}[2][]{\url{#2}}

\bibitem{Zwierz2010}
Zwierz M, P\'{e}rez-Delgado C~A and Kok P 2010 {\em Phys. Rev. Lett.\/} {\bf
  105} 180402

\bibitem{Giovanetti2012}
Giovannetti V and Maccone L 2012 {\em Phys. Rev. Lett.\/} {\bf 108}(21) 210404

\bibitem{Bollinger1996}
Bollinger J~J, Itano W~M, Wineland D~J and Heinzen D~J 1996 {\em Phys. Rev.
  A\/} {\bf 54} R4649--R4652

\bibitem{Lee2002}
Lee H, Kok P and Dowling J~P 2002 {\em Journal of Modern Optics\/} {\bf 49}
  2325--2338

\bibitem{Giovannetti2006}
Giovannetti V, Lloyd S and Maccone L 2006 {\em Phys. Rev. Lett.\/} {\bf 96}(1)
  010401

\bibitem{Huelga1997}
Huelga S~F, Macchiavello C, Pellizzari T, Ekert A~K, Plenio M~B and Cirac J~I
  1997 {\em Phys. Rev. Lett.\/} {\bf 79} 3865--3868

\bibitem{Escher2011}
Escher B~M, de~Matos~Filho R~L and Davidovich L 2011 {\em Nature Phys.\/} {\bf
  7} 406--411

\bibitem{Demkowicz2012}
Demkowicz-Dobrza\'{n}ski R, Ko\l{}ody\'{n}ski J and {Gu{\c t}{\u a}} M 2012
  {\em Nat. Commun.\/} {\bf 3} 1063

\bibitem{Helstrom1976}
Helstrom C~W 1976 {\em Quantum detection and estimation theory\/} (Academic
  press)

\bibitem{Holevo1982}
Holevo A~S 1982 {\em Probabilistic and Statistical Aspects of Quantum Theory\/}
  (North Holland, Amsterdam)

\bibitem{Anisimov2010}
Anisimov P~M, Raterman G~M, Chiruvelli A, Plick W~N, Huver S~D, Lee H and
  Dowling J~P 2010 {\em Phys. Rev. Lett.\/} {\bf 104}(10) 103602

\bibitem{Rivas2012}
Rivas A and Luis A 2012 {\em New J. Phys.\/} {\bf 14} 093052

\bibitem{Zhang2013}
Zhang Y~R, Jin G~R, Cao J~P, Liu W~P and Fan H 2013 {\em J. Phys. A\/} {\bf 46}
  035302

\bibitem{Berry2000}
Berry D~W and Wiseman H~M 2000 {\em Phys. Rev. Lett.\/} {\bf 85} 5098--5101

\bibitem{Kolodynski2010}
Ko\l{}ody\'{n}ski J and Demkowicz-Dobrza\'{n}ski R 2010 {\em Phys. Rev. A\/}
  {\bf 82}(5) 053804

\bibitem{Tsang2012}
Tsang M 2012 {\em Phys. Rev. Lett.\/} {\bf 108}(23) 230401

\bibitem{Nair2012}
Nair R 2012 {\em ArXiv e-prints\/} (\textit{Preprint} \eprint{1204.3761})

\bibitem{Hall2012}
Hall M~J~W and Wiseman H~M 2012 {\em New J. Phys.\/} {\bf 14} 033040

\bibitem{Giovannetti2012}
Giovannetti V, Lloyd S and Maccone L 2012 {\em Phys. Rev. Lett.\/} {\bf
  108}(26) 260405

\bibitem{Braunstein1994}
Braunstein S~L and Caves C~M 1994 {\em Phys. Rev. Lett.\/} {\bf 72} 3439--3443

\bibitem{Demkowicz2009}
Demkowicz-Dobrzanski R, Dorner U, Smith B~J, Lundeen J~S, Wasilewski W,
  Banaszek K and Walmsley I~A 2009 {\em Phys. Rev. A\/} {\bf 80}(1) 013825

\bibitem{Jarzyna2013}
Jarzyna M and Demkowicz-Dobrza\'{n}ski R 2013 {\em Phys. Rev. Lett.\/} {\bf
  110} 240405

\bibitem{Macieszczak2013}
Macieszczak K, Fraas M and Demkowicz-Dobrzañski R 2014 {\em New Journal of
  Physics\/} {\bf 16} 113002

\bibitem{Macieszczak2013a}
{Macieszczak} K 2013 {\em ArXiv e-prints\/} (\textit{Preprint}
  \eprint{1312.1356})

\bibitem{Frowis2014}
{Fr{\"o}wis} F, {Skotiniotis} M, {Kraus} B and {D{\"u}r} W 2014 {\em ArXiv
  e-prints\/} (\textit{Preprint} \eprint{1402.6946})

\bibitem{Knysh2014}
Knysh S~I, Chen E~H and Durkin G~A 2014 {\em ArXiv e-prints\/}  arXiv:1402.0495

\bibitem{Fujiwara2008}
Fujiwara A and Imai H 2008 {\em J. Phys. A: Math. Theor.\/} {\bf 41} 255304

\bibitem{Matsumoto2010}
Matsumoto K 2010 {\em ArXiv e-prints\/}  1006.0300v1

\bibitem{Kolodynski2013}
Ko{\l}ody{\'n}ski J and Demkowicz-Dobrza{\'n}ski R 2013 {\em New J. Phys.\/}
  {\bf 15} 073043

\bibitem{BrandorffNielsen2000}
Barndorff-Nielsen O~E and Gill R~D 2000 {\em J. Phys. A\/} {\bf 33} 4481--4490

\bibitem{Pezze2008}
Pezz\'e L and Smerzi A 2008 {\em Phys. Rev. Lett.\/} {\bf 100} 073601

\bibitem{Buzek1999}
Bu\v{z}ek V, Derka R and Massar S 1999 {\em Phys. Rev. Lett.\/} {\bf 82}(10)
  2207--2210

\bibitem{Berry09}
Berry D~W, Higgins B~L, Bartlett S~D, Mitchell M~W, Pryde G~J and Wiseman H~M
  2009 {\em Phys. Rev. A\/} {\bf 80} 052114

\bibitem{Demkowicz2011}
Demkowicz-Dobrza{\'n}ski R 2011 {\em Phys. Rev. A\/} {\bf 83} 061802

\bibitem{Gill1995}
Gill R~D and Levit B~Y 1995 {\em Bernoulli\/} {\bf 1} 59--79

\bibitem{Gaiba2009}
Gaiba R and Paris M~G 2009 {\em Physics Letters A\/} {\bf 373} 934--939

\bibitem{Vaart1998}
van~der Vaart A~W 1998 {\em Asymptotic Statistics\/} (Cambridge Univeristy
  Press)

\bibitem{Knysh2011}
Knysh S, Smelyanskiy V~N and Durkin G~A 2011 {\em Phys. Rev. A\/} {\bf 83}(2)
  021804

\bibitem{Gill2005}
{Gill} R~D 2005 {\em ArXiv Mathematics e-prints\/} (\textit{Preprint}
  \eprint{math/0512443})

\bibitem{Gill2013}
Gill R~D and {Gu{\c t}{\u a}} M 2013 {\em IMS Collections\/} {\bf 9} 105--127

\bibitem{Guta2007}
{Gu{\c t}{\u a}} M and Jen\u{c}ov\'{a} A 2007 {\em Commun. Math. Phys.\/} {\bf
  276} 341--379 ISSN 0010-3616

\bibitem{Kahn2009}
Kahn J and {Gu{\c t}{\u a}} M 2009 {\em Commun. Math. Phys.\/} {\bf 289}
  597--652

\bibitem{Chaves2013}
Chaves R, Brask J~B, Markiewicz M, Ko\l{}ody\'{n}ski J and Acin A 2013 {\em
  Phys. Rev. Lett.\/} {\bf 111}(12) 120401

\bibitem{Chin2012}
Chin A~W, Huelga S~F and Plenio M~B 2012 {\em Phys. Rev. Lett.\/} {\bf 109}(23)
  233601

\bibitem{LIGO2013}
LIGO{\;}Collaboration 2013 {\em Nature Photon.\/} {\bf 7} 613--619

\bibitem{Demkowicz2013}
Demkowicz-Dobrza\ifmmode~\acute{n}\else \'{n}\fi{}ski R, Banaszek K and
  Schnabel R 2013 {\em Phys. Rev. A\/} {\bf 88}(4) 041802

\bibitem{Bartlett2007}
Bartlett S~D, Rudolph T and Spekkens R~W 2007 {\em Rev. Mod. Phys.\/} {\bf 79}
  555

\bibitem{Jarzyna2012}
Jarzyna M and Demkowicz-Dobrza\'{n}ski R 2012 {\em Phys. Rev. A\/} {\bf 85}
  011801(R)

\bibitem{Demkowicz2014}
{Demkowicz-Dobrzanski} R, {Jarzyna} M and {Kolodynski} J 2014 {\em ArXiv
  e-prints\/} (\textit{Preprint} \eprint{1405.7703})

\bibitem{Chiribella2005}
Chiribella G, D'Ariano G~M and Sacchi M~F 2005 {\em Phys. Rev. A\/} {\bf 72}
  042338

\end{thebibliography}
\providecommand{\newblock}{}

\end{document}